# EFFECTS OF SHEAR-INDUCED CRYSTALLIZATION ON THE RHEOLOGY AND AGEING OF HARD SPHERE GLASSES

N. KOUMAKIS[1,2], A.B. SCHOFIELD[3] AND G. PETEKIDIS[1,2]

[1]FORTH-IESL, Road to Voutes, 71110, Heraklion, Crete, Greece

[2] Department of Materials Science and Technology, University of Crete, 71110, Heraklion, Crete, Greece

[3]School of Physics, The University of Edinburgh, Edinburgh, Kings Buildings, Mayfield Road, EH9 3JZ, U.K.

ABSTRACT: The rheological properties of highly concentrated suspensions of hard-sphere particles are studied with particular reference to the rheological response of shear induced crystals. Using practically monodisperse hard spheres, we prepare shear induced crystals under oscillatory shear and examine their linear and non-linear mechanical response in comparison with their glassy counterparts at the same volume fraction. It is evident, that shear-induced crystallization causes a significant drop in the elastic and viscous moduli due to structural rearrangements that ease flow. For the same reason the critical (peak of G'') and crossover (overlap of G' and G'') strain are smaller in the crystal compared to the glass at the same volume fraction. When, however the distance from the maximum packing in each state is taken into account the elastic modulus of the crystal is found to be larger than the glass at the same free volume suggesting a strengthened material due to long range order. Finally, shear induced crystals counter-intuitively exhibit similar rheological ageing to the glass (with a logarithmic increase of G'), indicating that the shear induced structure is not at thermodynamic equilibrium.

# INTRODUCTION

A large part of the interest in model colloidal systems such as hard spheres comes from their similarities to atomic systems, their ease of manipulation and the availability of experimental techniques that are able to probe them. Over the past few years, there has been a great amount of interest in shear induced ordering of colloidal suspensions[1-3]. Added to the general interest surrounding colloids, flow-induced ordering in colloidal systems have an added technological interest for the production of nanostructured materials for photonic[4], phononic[5], optofluidic[6] and other applications[7].

When left at rest, hard spheres exhibit a liquid phase at volume fractions below 0.494, a liquid-crystal coexistence phase at 0.494-0.545, and fully crystalline structure up to 0.58 where a kinetically frustrated glass state sets in[8]. The crystal structures to which hard spheres assemble when left at rest are a mixture of face-centred cubic (fcc) and hexagonally close packed (hcp) regions that are randomly orientated[9] which may age with time into a pure fcc crystal[10, 11]. The glass is a physically arrested state where each particle is trapped in a cage formed by its neighbours[8, 12]. In this metastable state, the system is far from the energetically preferred crystal structure and slowly explores the energetic landscape finding new minima with time[13, 14] a process that is manifested as ageing with a characteristic slowing down of the dynamics with waiting time[15]. Linear viscoelastic measurements in soft matter systems such charged spheres[16] reveal logarithmic ageing, while in others such as Laponites show an exponential one[17] In contrast, there seems to be no previous studies of the rheological ageing of hard sphere glasses and colloidal crystals.

Beyond the simplest system of hard spheres, shear induced ordering has been witnessed in a variety of systems such as charged colloids[18, 19] and microgels[20, 21]. Experimental techniques utilized to probe these systems include light[20, 22], X-ray[18, 23] and neutron scattering[21] coupled with rheology, while the viscoelastic properties of ordered charged colloids has been studied significantly[18, 19, 24, 25]. Furthermore, simulations of colloidal crystallization have been preformed for liquid[26, 27] and glassy systems[28]. The review of Vermant and Solomon[29] summarizes and gives an overview of the more important work.

More specific to this work, there have been direct observations of shear-induced crystallization of hard spheres with microscopy[3, 30] and light scattering experiments[1, 31]. Concentrated hard sphere systems crystallize under oscillatory shear with a large enough strain amplitude; in the liquid phase the crystal dissolves fully after cessation of shear whereas in the glassy state, shear-induced crystals are stable when the shear is turned off[31]. With respect to ageing glassy colloids are expected to exhibit three types of response: a) Very low strains in the linear regime do not perturb the internal mechanics and do not change the way the sample ages at rest, b) Mid range strains above the linear regime may induce over-ageing and c) High strains are able to slow down ageing (under-ageing) or even cause complete rejuvenation. This behaviour has been detected in glassy systems of charged spheres[32] and predicted by molecular simulations[14]. Although over-ageing is not always observed[33], it widely acceptable that high strains prevent ageing. However, it should also be kept in mind that whether high shear may literally rejuvenate a glassy material is also under debate [34].

Even though this subject has been examined for many years, a complete study of the rheology of the shear-induced crystal structures in comparison to that of the hard sphere glass is still lacking. Here we present rheological measurements on hard sphere glasses and crystals produced when the former are submitted to large amplitude oscillatory shear. The linear and nonlinear rheological response of the glass and crystal are probed by dynamic frequency and strain sweeps and are compared against each other at various volume fractions. Moreover, we monitor the rheological ageing of both states, i.e. the evolution of their linear viscoelastic properties with elapsed time after rejuvenation. The rest of the paper is organised as follows: We first introduce the experimental details regarding samples and measurements. Afterwards, we present and discuss our experimental findings on the crystal creation, linear viscoelasticity, details of nonlinear measurements and the ageing behaviour; finally we close with the conclusions.

## EXPERIMENTAL DETAILS

### Samples

The colloidal particles consisted of polymethylmethacrylate (PMMA) particles sterically stabilized by a thin chemically grafted layer of poly-12-hydroxystearic acid chains suspended in cis-decahydronaphthalene (cis-decalin) were they have been shown to interact as hard spheres[35]. For long time ageing experiments, particles were dissolved in octadecene to avoid solvent evaporation. The radii, determined by light scattering, were R=267nm in cis-decalin and R=288nm in octadecene. In addition, few measurements were conducted with larger particles (R=689nm) in cis-decalin. Both particle dispersions had low enough polydispersities (~5%) allowing them to crystallize at rest and under shear. The volume fraction for samples in octadecene and for the larger spheres in cis-decalin where crystallization was difficult to observe, where determined from random closed packing (prepared by centrifugation) set to 0.66, according to computer simulations[36]. The volume fraction of the smaller particles in cis-decalin was determined in the co-existence region, giving a close packing volume fraction, produced by centrifugation, of 0.673. This difference can be attributed to partial ordering during centrifugation of smaller particles in cis-decalin as the particle density mismatch and thus the sedimentation speed is lower[37]. The rest of the concentrations were determined by successive dilutions of the same sample batch. Before measurements the samples were thoroughly mixed and experiments started immediately after loading following the rejuvenation protocol described below.

## Rheology

A Rheometric Scientific stress controlled DSR Rheometer was used for all measurements involving optical observations. At the expense of optical feedback, a Rheometric Scientific strain controlled ARES Rheometer was also used for any situation that required constant strain or Peltier temperature stabilization. All the measurements were made at a constant temperature of 20°C and loading history was erased before starting measurements by applying a low rate steady shear for about 10 seconds so as to destroy any crystallization induced by loading.

Shear-induced crystallization during rheological measurements was followed monitoring Bragg scattering from a laser beam impinging on the sample through transparent rheometer tools. Measurements using parallel glass plates revealed partial crystallization of the sample due to inhomogeneous strain and consequently problematic interpretation of the data. To overcome this problem, a transparent Plexiglas cone with a diameter of 38mm and an angle of 0.03 rad was constructed.

The scattered pattern was used to determine the amount of crystallization as well as the type and orientation of the crystal structure. An amorphous glass state produced a Debye-Scherrer ring whereas a crystallized sample was revealed through Bragg peaks; their orientation relative to the shear direction was determined by the crystal orientation. When the sample was semi-crystalline, a mixture of both scattering patterns could be seen.

Before the crystal was created, a Dynamic Frequency Sweep (DFS) in the linear regime was performed on the glass in order to determine the elastic (G') and viscous (G'') moduli before the onset of ageing. In order to create the crystal, a Dynamic Stress Sweep (DSS) of increasing stress was performed until the sample was fully crystallized. Immediately after crystal creation, the samples were probed with a DSS of decreasing stress and subsequently with a DFS in the linear regime so as to measure the linear viscoelastic properties of the shear induced crystal. Afterwards a DSS of increasing stress was also performed on the crystal to investigate possible hysteresis phenomena. For the ageing studies, the sample was probed in the linear regime for a short time every 30 minutes in order to minimize stress induced ageing. Recent work[38] showed that slip and shear banding may be observed in shear induced crystals under oscillatory shear while steady shear microscopy measurements[39] have also given some indication of shear banding. Here, direct observation of the sample in the cone-plate geometry using a CCD camera and a magnifying lens revealed no evident slip or clear shear banding; however due to the crudeness of the method the later can not be totally excluded. On the other hand, confocal imageing of similar fluorescent PMMA particles in a refractive index matching solvent (decalin/tetralin mixture) has shown that particles adjacent to the glass plates were mobile whereas in slightly mismatched samples (decalin) they appear to stick on them due to van der Waals attractions[40]. These observations corroborate with slip under steady shear seen in the former case[41] and its absence in similar measurements of non refractive-index matched samples[42, 43].

## RESULTS – DISCUSSION

### Crystal Creation – Dynamic strain sweeps

Figures 1 and 2 show the DSS measurements of four different volume fractions in cis-decalin in the glass regime ($\varphi$=0.610, 0.619, 0.641 and 0.656). The measurements were conducted at a frequency of 10rad/sec and plotted as a function of strain instead of stress, since yield strain is expected to depend on volume fraction much less than yield stress[42]. The linear regime, where the solid like behaviour (regime a) of the glass is demonstrated by a frequency independent G' an order of magnitude larger than the G'', extended to about 1% strain amplitude. Above this value shear thinning sets in with G' decreasing and G'' increasing with strain amplitude. At about 10-20% strain, G'' crosses over G' (regime b) which is characteristic of a transition from viscoelastic solid- to liquid-like behaviour and thus provides a measurement of the yield strain. Moreover, in this region Bragg spots with a characteristic six-fold pattern began to appear, indicating the onset of crystal creation. At this point, the Bragg spots were still faint and the underlying image of a diffuse Debye-Scherrer ring was dominant. As the strain was increased, the amorphous ring progressively disappeared as it was replaced by more intense Bragg peaks. At a strain of about 100%, full crystallization was achieved (region c) with only the high intensity Bragg peaks remaining. The speed of crystal creation and its stabilization was dependent on the amount of shearing time, so in order to approach equilibrium at each strain measured the number of strain points and the shearing time of each point were large (typically 30 points per decade at 30 cycles per point). Applying higher strains (>150%) led to disruption of the crystal structure and reappearance of the amorphous ring (regime d). Measurements at high volume fractions would additionally exhibit shear thickening at these high strain amplitudes leading to crystal melting similar to steady shear experiments[44]. Thickening was more evident for the large spheres and high frequencies, as expected[45], and was systematically avoided hence not shown in any of the figures.

Figures 1 and 2 also show DSS runs at decreasing strain where the viscoelastic properties of the shear induced crystal structure were probed. If upon increasing the strain, the crystal was breaking when high strains were reached, on decreasing the strain, the crystal was reformed as verified by the intensity of the Bragg peaks. This reformation is the origin of the sharp drop of G' and G'' when reducing strain seen in Figure 1a and 1b. If however the crystal did not break during strain increase, the backward sweep probed the crystal structure in reverse as shown in figures 2a and 2b. In all these measurements, the most obvious and interesting finding in the linear regime was the drop of both G' and G'' when the glass was converted through shear into crystal. Such a drop of the viscoelastic moduli, which could be of more than one order of magnitude, was found to increase with volume fraction. Performing a second increasing DSS revealed little hysteresis in the mechanical response of the crystal structure (see Figure 1). At the point where G' and G'' of the first (glass) and third (crystal) forward DSS begin to merge (f), the crystal structure was optically observed to start breaking. Furthermore, a mixture of Bragg peaks and an amorphous ring was seen until the strain reached point (d) where the crystal dissolved completely and the Bragg peaks disappeared.

The sixfold pattern of the Bragg peaks observed in figure 1a suggests that a random stacking of layers was formed with the (111) plane parallel to the rheometer plates, analogous to previous findings in suspensions of hard spheres[1, 3] and microgel particles[20]. However, in the work of Haw et al.[3], low oscillatory strains (<50%) produced fcc crystallites with a preferred close packed direction perpendicular to shear, while at high strains (>50%) random hexagonal layering was observed with a close packed direction parallel to shear in qualitative agreement with the previous experiments by Ackerson[1]. Here instead we only distinguished Bragg patterns corresponding to random layering with close packed direction parallel to shear. The discrepancy probably emanates from the specific rotational cone-plate geometry where the crystallites are constrained in such a way that promotes growth only parallel to shear. This was verified by conducting oscillatory tests in a sliding parallel plate shear cell[46] where crystals with close packing direction both perpendicular (at low strains) and parallel (at high strains) to shear were observed.

It is reasonable to expect that shear induced crystallization takes place because a crystal may be strained more easily and exhibits less frequent particle collisions than the same volume fraction glass[1, 30]. This could be the origin of the significantly lower viscoelastic moduli of the crystal since the material assembles into the crystal structure in an effort to ease the imposed stress. It further corroborates with the fixed orientation and monocrystallinity of shear induced crystals, as opposed to the polycrystalline structures formed at rest.

Alternatively, shear-induced crystallization can also be described from an energetic point of view. Since the crystal is the equilibrium phase, while the glass is a kinetically frustrated one, shear may be considered as a mechanism that provides the energy needed for the system to crystallize. With increasing strain, entropic barriers trapping particles in cages are slowly reduced and disappear[28] resulting in an increased out of cage particle diffusion that allows the particles to rearrange into an energetically preferred crystal structure. Recent Brownian dynamics simulations and oscillatory DWS echo experiments[47] indicate such increased shear induced diffusion. Even when the barrier is reduced but not eliminated, the particles are given an increased possibility of escape from the cage that might lead to partial crystallization. With increasing oscillation frequency, the particles are given more opportunities to escape, which means that the crystallization process will be more rapid, agreeing with observations of shear induced crystallization in colloid-polymer gels[48].

In all our experiments glasses start crystallizing at a strain of about 10%-20%, around the crossover point of G' and G''. This value is very close to the yield strain of the polydisperse hard sphere glasses[42, 49] and could further be identified as the shear induced analogue of the Lindermann criterion[50] for hard sphere freezing. While lower strains could not induce any crystallization even after long times (1 hour), at 100% strain the glass would fully crystallize quite fast (about one minute). Moreover, for the larger spheres shear induced crystallization progressed considerably faster; evidently due to the reduced contribution of Brownian motion. These findings suggest that local order is promoted by repeated direct near neighbour interactions when a) the yield strain is exceeded and b) Brownian diffusion is not fast enough to remix particles within one period of oscillation. This is the reason why DSS at l rad/sec was not able to induce full crystallization in small particles. Thus all DSS experiments were done at a frequency of 10 rad/sec in which both spheres could crystallize easily.

For low shear rates, $\dot{\gamma}$ (=$\omega\gamma_0$), such that the Peclet number, Pe=$\dot{\gamma}\tau_B$<<1, the intrinsic relaxation in the system, $\tau_B$, is faster than the rate at which shear disturbs the structure and thus no crystallization is expected even if the strain amplitude is larger than the yield strain. If however the shear rate is sufficiently high, (Pe>1) and the strain amplitude larger than the yield strain the structure of the system may be altered before it can relax back to equilibrium. While for dilute suspensions the diffusion time over a distance equal to the particle radius, R, is $\tau_B=R^2/6D$, with D the Stokes-Einstein-Sutherland diffusivity (=$k_BT/6\pi\eta R$) in a medium of viscosity η, for highly concentrated and glassy states the long-time out of cage diffusion is very slow or completely frozen. According to a dynamic criterion at freezing the long–time diffusion is close to 10% of the short-time one[51]. Relating the findings here with the above criterion and in accordance to observations in crystallizing colloid-polymer gels[48] we may reasonably argue that crystallization under oscillatory shear would take place when the long time shear induced diffusion becomes faster than 10% of the short-time one. Hence, assuming that $1/\tau_{long}(\dot\gamma) \propto 1/\tau_{long}(\dot\gamma=0)+\dot\gamma$ and for $\tau_{long}(\dot\gamma=0) \gg \tau_B$ and a characteristic time at crystallization, $\tau_{long}(\dot\gamma=\dot\gamma_{cr})=10\tau_B$ we get: $\tau_B\dot\gamma_{cr}=0.1$. For the present system of the smaller particles in cis-decalin at $\gamma_0$=20% (near the yield strain, where the sample crystallizes) and ω=10rad/sec we calculate $\tau_B\dot\gamma=0.093$, close enough to the predicted value to support the above, simple, argumentation.

The crystal created could be dissolved either by applying an oscillatory shear with a high enough strain, or by applying a steady shear. Large oscillatory strains (>150%) led either to shear thickening (for larger particles) or simply to the breaking of the crystal structure. As expected, smaller particles did not shear thicken easily except at the very high volume fractions and strains (>250%) or frequencies. In order to avoid complications stemming from shear thickening whenever there was need to go over from crystal to glass, a low rate steady shear was applied.

**Linear viscoelasticity of glass and crystal**

In Figures 3 and 4 we show the linear viscoelastic data of the glass and shear induced crystal at different volume fractions in cis-decalin. The dynamic frequency sweeps were performed in the linear regime with a strain of 0.5%. The frequency dependence of crystal and glass are similar, while G' and G'' of the crystal were about one order of magnitude lower than those of the glass. Both for the glass and the crystal and for all volume fractions, G' exhibited a slight increase with frequency and in most cases G'' showed a minimum while the crystal systematically had a slightly larger slope of both G' and G'' with frequency. For the frequency range and volume fractions measured here G'' did not show a clear high frequency, $\omega^{1/2}$, behavior observed in similar systems at lower volume fractions[52, 53] due to the Brownian contribution. Moreover, the minimum of G'' shifts to lower frequencies when a glassy sample crystallises under shear, as well as when the volume fraction decreases (see figs. 3 and 4).

The fact that G'' rises at low frequencies indicates the existence of a slow dissipative process which is not anticipated by "ideal" glass models such as mode coupling theory. Nevertheless, such additional slow relaxation modes have been seen in a wide range of soft matter glassy systems by dynamic light scattering[15]. On the other hand, in the viscoelastic spectra the time scale corresponding to the G'' minimum describes the transition from a relaxation mode related to the fast in cage diffusion, to a slower, long-distance, out of cage motion. The latter has been described also in terms of hopping mechanisms thermally activated at rest[54]. With this viewpoint fittings according to mode coupling theory predictions for the linear viscoelasticity of concentrated hard sphere suspensions[52] were included in figures 3 and 4. Assuming that the stress and density autocorrelation functions (the latter connected to the intermediate scattering function in light scattering) have the same form, the frequency dependence of the storage and loss moduli on the liquid side were calculated[52]:

$$G'(\omega) = G_P + G_\sigma \left[ \Gamma(1-a')\cos\left(\frac{\pi a'}{2}\right)(\omega t_\sigma)^{a'} - B\Gamma(1+b')\cos\left(\frac{\pi b'}{2}\right)(\omega t_\sigma)^{-b'} \right]$$

$$G''(\omega) = G_\sigma \left[ \Gamma(1-a')\sin\left(\frac{\pi a'}{2}\right)(\omega t_\sigma)^{a'} + B\Gamma(1+b')\sin\left(\frac{\pi b'}{2}\right)(\omega t_\sigma)^{-b'} \right] + \eta'_\infty \omega$$

Guided by our experimental data we have omitted here the Brownian contribution term $\sim \omega^{1/2}$ at high frequencies as its addition contributed less than 1% in the frequency range measured. Here $\Gamma(x)$ is the gamma function, a'=0.301, B=0.963 and b'=0.545 are parameters predicted for hard spheres[55], $\eta'_\infty$ is the high frequency viscosity, $G_\sigma$ the viscoelastic amplitude that determines the variation of G'($\omega$) and the magnitude of G'' at the minimum, while $G_P$ is the plateau value of the elastic modulus. The time $t_\sigma$ corresponds to the inverse frequency where the minimum of G'' occurs and according to MCT represents the crossover from the β to the α process rather than a characteristic relaxation time[56], where the β process is related to the short time, in-cage relaxation and the α process to the long time, out of cage relaxation. In the 'ideal' glass however, the α process should be frozen and thus $t_\sigma$ should reflect the crossover towards the non-ergodicity plateau. Nevertheless, in real systems where ultra-slow ageing modes are observed, $t_\sigma$ and thus the minimum of G'' describe the transition to the ultraslow relaxation.

With the assumption that the MCT model for the linear rheology might be used inside the glass state, the fits in figures 3 and 4 primarily are used to extract the characteristic time related with the G'' minimum. Incidentally, it is interesting to note that the crystal data can also be fitted reasonably well by the MCT approach. This approximation might still hold since the 'ideal' α relaxation of MCT, active on the liquid side, is replaced by an ultraslow, ageing, mode which is the origin of the energy dissipation at low frequencies in the glassy state.

Figure 5a shows $t_\sigma$ from the MCT fits as a function of the volume fraction for the glass and shear induced crystal. In both cases, $t_\sigma$, decreases with increasing volume fraction, with the time deduced from the crystal being larger than that of the glass. Although the minimum is not apparent in the experimental data for lower volume fractions, using the G'' minimum as a free fit parameter for the theory we find an increasing time scale as the volume fraction decreases. The values of $G_p$ and $G_\sigma$ used in the MCT fits are shown in fig. 5b. Both increase with volume fraction as expected from figures 3 and 4, while those for the crystal are lower than for the glass.

We believe that $t_\sigma$ is coupled with the average time a particle needs to explore its surrounding cage. Thus, when volume fraction increases, particle cages become tighter and the characteristic time decreases. Similarly in shear-induced crystals, the average interparticle distance increases compared to the glass of the same volume fraction and the characteristic time is thus larger. Assuming that the short-time diffusion coefficient corresponding to motions within the cage is the same in the glass and crystal we can estimate that $t_{crystal} = \frac{\Delta_{crystal}^2}{\Delta_{glass}^2} t_{glass}$ where $\Delta = 2R((\phi_m/\phi)^{1/3} - 1)$ is the average distance between particles and $\varphi_m$ the maximum packing fraction (0.66 for the glass and 0.74 for the crystal). The solid line shown in fig. 5a depicts the estimated time for the crystal using the glass data according to the above. The relatively good agreement with the experimental data from the shear induced crystal supports such rationalization.

Furthermore, MCT predicts that the short time dynamics, as measured for example by dynamic light scattering, are similar for the same distance from the glass transition volume fraction, $\varphi_g$, in the fluid and glass side[57]. This behaviour, represented by a sharp peak of the β relaxation time, $\tau_\beta$, around $\varphi_g$[56], suggests that as the volume fraction is increased inside the glass state the minimum of G'' would shift towards higher frequencies in agreement with our experimental findings. Note, however, that the MCT predictions do not take into account hydrodynamic interactions which slow down particle motion at high φ. Nevertheless, a speed up of $t_\sigma$ below φ=0.58 was not evident in our samples probably due to low torque conditions hampering accurate measurements of G''.

**Elastic modulus of glass and crystal: Volume fraction dependence**

Figure 6 shows the elastic modulus, G', of the glass and shear induced crystal against volume fraction for samples in cis-decalin and octadecene. G' values were taken at a frequency of 1 rad/sec and normalized with sphere size and thermal energy. A reasonably good agreement of G' data is observed in the two solvents for both the glass and the crystal. As shown in figs. 3 and 4 and in agreement with $G_p$ and $G_\sigma$ of the MCT analysis (fig. 5b), G' of the crystal acquires both smaller absolute values and a weaker volume fraction dependence compared to the glass. The basic finding that G' of the crystal is lower than that of the glass at the same volume fraction is in accordance with density functional calculations[58] where the shear modulus of a glass with random closing packing of 0.66 was found to be larger than that of an fcc crystal for all shear directions.

Figure 6 includes predictions for the elastic modulus in a hard sphere fcc crystal according to a weighted-density-functional theory[59] and molecular dynamics simulations[60] for the (111) plane in the velocity-vorticity plane and the close packed direction parallel to shear. In both cases the theoretical curves do not agree well with the experimental ones; however they are of the same order of magnitude for volume fractions around 60%. It is interesting to note that the prediction of the activated hoping MCT model[54] for an exponential volume fraction dependence in a hard sphere glass yields larger values than the crystal but with similar slope, whereas the experimental data show a steeper increase (fig. 6). Thus, it seems that an accurate theoretical description of the elastic modulus of hard sphere glasses is still lacking.

However, it is still intriguing that the shear induced crystal has weaker elastic and viscous moduli than the glass, since intuitively it might be expected that stronger long range order would promote solid like behavior and increase the elastic modulus. We believe though that the determining quantity is not the absolute volume fraction which in both states is the same, but rather the distance from the maximum close packing which is different. To investigate this dependence we plotted G' and G'' for both systems as a function of the average free volume available around a particle. Hence, figure 7 shows the viscoelastic moduli as a function of the inverse distance of the volume fraction from the maximum packing fraction $1/\varphi_{free}=1/(\varphi_m-\varphi)$. For a glass, the maximum packing fraction is the random close packing of $\varphi_m=0.66$ whereas for the crystal 0.74 is the fcc maximum packing. Then the situation is qualitatively reversed: figure 7 reveals that G' of the crystal is larger than that of the glass for the same distance from maximum packing. As stated before this is to be expected when comparing an ordered structure with an amorphous one, nevertheless these experiments clarify that such intuitive expectation holds only at the same average free volume rather than at the absolute one. Along the same line we prepared a polycrystalline sample at rest, inside the plates of the rheometer, at φ=0.55. Linear viscoelastic measurements of such polycrystalline sample yielded a higher G' compared to the shear aligned one. This finding reflects the different values of the elastic modulus in different shear directions as predicted by simulations and theory[59, 60]. Polycrystalline samples with random orientation of crystallites have higher G' than those aligned or formed by shear but lower than a completely amorphous glass sample.

A way of characterising inter-particle interactions is to fit the G' data with a power law[20] of the form $G' \sim \varphi^m$. Measurements of the glass samples in cis-decalin yield an exponent of $m_{dec}$=48 while in octadecene $m_{oct}$=42 in agreement with our previous experiments with similar PMMA particles[43]. Note however that these exponents are slightly dependent on the frequency. The fitting parameter $G_p$ (figure 5) follows similar strong volume fraction dependence. Systems with softer interactions are known to exhibit weaker exponents in the range of m=4-7[61, 62]. On the other hand, as stated above, the volume fraction dependence of G' (and G'') for the crystal is weaker than that of the glass (figure 6). For both samples the power law exponent for the crystal is approximately half of that in the glass. They coincide however if the frequency at which the glass sample is probed is around 200 rad/sec (keeping that for the crystal at 1 rad/sec). The high frequency modulus $G_\infty$, which is directly related to the interparticle potential[20], should be identical in both crystal and glass. To this end, it might be interesting to note that crystals and glasses display similar high frequency phonon dispersions, as measured by Brillouin light scattering[63] suggesting that although the macroscopic elastic low frequency response is clearly affected by long range order, the high frequency one seems to be dominated only by the local order at the level of the first neighbours.

**Crystal and Glass Critical Strain**

Figure 8 shows the volume fraction dependence of the critical strain (determined at the maximum of G'') and the crossover strain (defined at the G'=G'' point) as deduced from the DSS data in cis-decalin at a frequency of 10rad/sec (figures 1 and 2). The critical strain is the point where the system manifests a maximum viscous response (G'') while the crossover strain is the point above which viscous behaviour dominates over elastic one and strong irreversible rearrangements begin to occur. The former signifies the point of maximum energy dissipation associated with structural changes while the latter often is used to define the yield strain.

Figures 1 and 2 reveal a broader G'' peak for the glass samples compared to that in the shear induced crystal. This effect might be related with the process of crystal formation during a progressive increase of the strain amplitude in a DSS experiment. In this sense, the broadness of the G'' peak reflects the existence of two interrelated, but not identical, dissipation mechanisms; crystallization and yielding. The latter naturally promotes the former, although partial crystallization would also facilitate easier flow in the sheared sample. Furthermore, in shear induced crystals the G'' peak coincides with the G'-G'' overlap whereas for glass samples in general the peak occurs at strains below the G'-G'' overlap supporting the idea of interconnected yielding and crystallization mechanisms. In comparison, experiments on polydisperse hard and soft sphere glasses, where no crystallization is observed under shear, show simpler dynamic strain sweeps where the peak of G'' superimposes with the G'-G'' overlap[61, 64].

The critical strain for both the glass and the crystal drops with increasing volume fraction as seen in Figure 8a. We may attribute such behaviour to the decreasing inter-particle distance with increasing volume fraction that leads to a lower strain at which particles would start colliding with each other, enhancing energy dissipation. The critical strain for the crystal does not drop as fast as that in the glass due to the larger distance from maximum crystal packing (0.74) as opposed to random close packing (0.66).

The idea that the maximum energy dissipation would be achieved at the strain amplitude where particles would start colliding with their neighbors may lead to simple calculation of a critical strain that decreases with increasing volume fraction and becomes zero at maximum packing. Calculating the volume fraction dependence of the average inter-particle distance yields a strain of $1-(\frac{\phi}{\phi_m})^{1/3}$, which apparently is rather low compared to experimental data [42].

However, a more careful calculation of the maximum strain that may be accommodated in a sheared suspension before a particle comes into contact with its surrounding cage (which under shear is forming an ellipsoid elongated along the shear direction) gives[65] $\gamma_{max} = \frac{4[(\frac{\phi_m}{\phi})^{2/3} - (\frac{\phi_m}{\phi})^{1/3}]}{2(\frac{\phi_m}{\phi})^{1/3} - 1}$. These simple calculations (figure 8a) suggest that shear induced particle collisions may well be the origin of the G'' peak in concentrated suspensions of hard spheres. Such collisions would also be related in the present system with shear induced crystallization.

The glass seems to generally have a higher crossover strain than the crystal (figure 8b), exhibiting a clear peak in agreement with our previous rheological and LS-echo measurements[42, 46] in non-crystallizing polydisperse hard sphere glasses. For a glass the crossover strain defines the onset of melting accompanied by large scale irreversible particle rearrangements and the 'breaking' of cages. Moreover, we have shown[49] that the yield strain determined from creep and recovery measurements, as the maximum recovered strain after stress removal, is also showing a maximum with volume fraction in agreement with the behaviour of the crossover strain observed here. For a crystal, the crossover strain is again accompanied by irreversible rearrangements and the point were the crystal flows due to slipping crystal layers[1]. The reason that the crystal generally has a lower crossover strain than the glass is probably because the slipping layers make it easier for a crystal to actually start flowing.

The maximum in the yield strain is the result of two competing mechanisms. In the glass state, as the volume fraction decreases, the cages loosen up and the yield strain starts decreasing until the system becomes a liquid where it goes to zero. On the other hand, as $\varphi$ increases towards maximum close packing, tighter cages are formed that lower the yield strain and thus make the material more brittle. Activated hopping MCT predictions for the yield strain of hard sphere glasses[54] qualitatively agree with our experimental data. The crossover strain for the shear induced crystal is less sensitive to volume fraction (figure 8b and 9b). This might be related to the increased free volume available for each particle due to larger distance from maximum packing or to distinct hydrodynamic interactions in the ordered system. The detailed physical mechanism of such a discrepancy, as with the other specific differences in the flow and yielding of the glass and the shear induced crystal, is still unclear and calls for a comprehensive theoretical description.

Figure 9 shows the crossover and critical strains as a function of the inverse free volume available. For both the glass and the crystal the critical strain drops with increasing volume fraction (decreasing $\varphi_{free}$), however, now it is clear that the drop is larger for the glass due to the closer proximity to random close packing. It is also worth noting that the crossover strain plotted against the inverse free volume exhibits, interestingly, a common power law increase (solid line in figure 8b) for both the glass and the crystal until the maximum yield strain. This might be an indication of a common mechanism underlying the initial increase of $\gamma_{cross}$.

## Ageing of the glass and crystal

Finally, we discuss the evolution of the viscoelastic moduli with waiting time (ageing). Figure 10 depicts the time dependence of G' for glass samples of different volume fractions together with those of the respective shear-induced crystals. In order to exclude evaporation and sedimentation effects during the long periods of measurements the smaller particles in octadecene were used. Zero time is defined by shear rejuvenation in the case of the glass and the time at which crystallization was fully evolved for shear induced crystals. Even though continuous measurements showed minimal effects, to avoid a possible stress induced ageing, each data point in figure 9 was measured for a relatively short time every 30 minutes.

The glass sample exhibits rheological ageing with increasing G' versus the elapsed time from rejuvenation. More specifically G' measured in the linear regime ($\gamma_0$=1%) at 10rad/s is found to increase logarithmically for about ten hours with strong fluctuations observed after that. Higher volume fractions show a stronger increase in G', with a larger slope in the semi-logarithmic plot. At the same time G'', after an initial drop as a result of rejuvenation (typically <100s), also increases with a smaller slope (not shown). Similar evolution of the linear viscoelastic properties with age have been observed in colloidal pastes[16], attractive colloidal gels[66] and industrial systems such as clays[67]. Although the detailed mechanism of ageing in glassy systems is currently the subject of a number of theoretical studies[13, 16, 68] the general physical picture arising is that of a progressive evolution of the system towards deeper minima of an energy landscape. This slow drift into deeper metastable states might be driven by thermal motion, mechanical perturbation or other external disturbances such as temperature fluctuations. In the metastable glass state particles do not have enough thermal energy to overcome the entropic barriers in order to evolve towards the energetically preferable crystal structure. Shearing causes energy barriers to drop allowing particles to rearrange into an ordered structure. Hence, hard sphere crystals might, ideally, be expected to show no ageing as opposed to the same volume fraction glass.

It should be mentioned that for shear-induced crystal left at rest for long periods of time (>10 hours) Bragg peaks seemed to be stable suggesting that once formed the crystal keeps its structural integrity and does not dissolve. However, when it was rheologically probed by small amplitude oscillations, a similar rheological ageing to that of the glass was observed (figure 10). This contradicts the existence of a thermodynamically stable crystal phase and implies that perhaps the shear-induced crystal is not identical to the equilibrium crystal at rest due to the rotational cone-plate geometry. Thus, after cessation of large amplitude oscillatory shear, we believe that the crystal slowly evolves towards the preferred non-rotational fcc structure, similarly to how hard spheres rearrange from random hcp crystallites to pure fcc at rest[11]. The latter does not necessarily have the closed packed direction aligned with shear and since theory[59] and simulations[60] suggest that other orientations have higher elastic constants, this could explain the observed increase with waiting time. In summary, it seems that crystal ages, but there still remains an open question on the effect of the geometry. To this end, simultaneous microscopy and rheology should be able to clarify this issue.

In addition, a correlation between small changes in measured temperature (± 0.1 ºC) and ageing was observed. Strong fluctuations of G' in both the glass and the crystal were correlated with small temperature fluctuations inducing stresses in the sample or shearing flow as observed recently in other soft matter glasses[69]. Such effects may result in over- or under-ageing of the sample. Experiments with a Peltier temperature stabilized system (± 0.01 ºC) instead of the standard bath minimized these fluctuations which however were still present somewhat less in frequency and magnitude supporting the idea that temperature fluctuations are not the cause of ageing itself but rather of the erratic G' and G'' fluctuations.

## CONCLUSIONS

We have shown that hard sphere colloidal glasses crystallize under oscillatory shear at strain amplitudes above the yield strain (about 10-15%). The viscoelastic moduli of the shear induced crystals were found to be significantly lower than those of the glass at the same volume fraction. We argue that this results from a mechanism that reduces stresses in the sheared material. The storage and loss moduli of the crystal exhibit a weaker increase with volume fraction compared to the glass. When, however, G' and G'' are plotted as a function of the inverse free volume, taking into account the distance from maximum packing in each state, the crystal is found to have a larger G' than the glass at the same free volume signifying the effects of long range order in strengthening the solid-like character of the sample.

The linear viscoelastic data of both the glass and crystal may be fitted quite well by MCT predictions, yielding in both states a characteristic crossover time (determined at the minimum of G'') that decreases as the volume fraction increase due to a progressively tighter cage. This crossover time is longer in the crystal due to a larger on average, free volume available to a particle compared to the glass of the same volume fraction. We also found that the crossover strain of the glass is generally higher than its crystal counterpart at the same volume fraction probably due to slipping layers that may allow easier yielding. Additionally, both the yield and crossover strain of the crystal structure are less affected by volume fraction compared to the glass due to larger distance from maximum crystal packing.

Finally, contrary to intuition, shear induced crystals seem to age rheologically much like the glass itself indicating that the initial crystal formed in the cone-plate geometry is not in a thermodynamic equilibrium.

**Acknowledgements:** We thank M. Cates, J. Mewis and J. Brady for fruitful discussions. This work was supported by the Greek General Secretariat for Research and Technology (Basic Research Program PENED,-03ED566) the EU ToK project COSINES and the EU Softcomp Network of Excellence.

**Figure captions:**

Figure 1: Dynamic strain sweeps of low volume fraction hard sphere glasses (with R=267nm) in cis-decalin at a frequency of 10rad/sec: a) φ=0.610, b) φ=0.619. G' is represented by solid symbols and G'' by open ones. Three runs are shown: Upward strain sweep in a glass sample (black symbols), downward (red) and upward (blue) strain sweeps of a crystal. The arrows indicate the direction of changing strain and the letters denote positions in the crystallization process discussed in the text. The scattering patterns for different stages of crystallization are shown in the inset photos of Figure 1a (the arrow shows the direction of shear).

Figure 2: Dynamic strain sweeps of high volume fraction hard sphere glasses (with R=267nm) in cis-decalin at a frequency of 10 rad/sec a) 0.641, b) 0.656. G' is represented by solid symbols and G'' by open ones. Two runs are shown: Upward strain sweep in a glass sample (black symbols) and downward (red) strain sweeps of a crystal. Arrows indicate the direction of changing strain. Vertical arrows denote the critical (peak of G'') and crossover/yielding (G'=G'') points.

Figures 3: Linear viscoelastic data at low volume fraction samples in cis-decalin: a) φ=0.600, b) 0.610 and c) 0.619 for the glass (black circles) and the crystal (red squares). G' is represented by solid symbols and G'' and by open ones. Added are fits to the data by the MCT theory[52].

Figure 4: Linear viscoelastic data at high volume fraction samples in cis-decalin: a) φ= 0.641 and b) 0.656 for the glass (black circles) and the crystal (red squares). G' is represented by solid symbols and G'' and by open ones. Added are fits to the data by the MCT theory[52].

Figure 5: a) Volume fraction dependence of the crossover time, $t_\sigma$, corresponding to the minimum of G'' (MCT fits of figures 3 and 4) for the glass (black circle) and shear induced

crystal (red square). The line depicts the prediction for the $t_\sigma$ of the crystal based on the glass data according to $\tau_{crystal} = \dfrac{\Delta^2_{crystal}}{\Delta^2_{glass}}\tau_{glass}$. b) Volume fraction dependence of MCT fitting parameters $G_p$ (solid symbols) and $G_\sigma$ (open symbols) for the glass (circles) and the crystal (squares).

Figure 6: Volume fraction dependence of the normalized G' for the glass and the shear induced crystals in cis-decalin and octadecene as indicated. The lines represent the corresponding predictions from simulations and density function theory for the crystal and the activated hopping MCT model for the glass. Crystal predictions correspond to the elastic constant $C'_{44}$ for shear with the velocity-vorticity plane parallel to (111), derived from the independent elastic constants of the primitive fcc cell according to[59, 60], $C'_{44}=(C_{11}-C_{12}+C_{44})/3$.

Figure 7: The normalized G' for the glass and the shear induced crystals in cis-decalin and octadecene as a function of the inverse free volume. Maximum packing is 0.66 (RCP) for the glass and 0.74 for the fcc crystal. The lines represent the corresponding predictions from simulations and density function theory for the crystal and the activated hopping MCT model for the glass as indicated.

Figure 8: Volume fraction dependence of a) the critical strain for the glass (solid circle) and the crystal (open circle) and b) the crossover strain for the glass (solid circle) and the crystal (open circle). The points were taken from DSS at $\omega=10$rad/sec as shown in figure 2. The dotted and solid lines in figure 8a represent the simple predictions for the maximum strain that may be accommodated before particles start hitting each other for a static and a sheared cage respectively, using $\varphi_m=0.66$.

Figure 9: a) The critical strain for the glass (solid circle) and the crystal (open circle) and b) the crossover strain for the glass (solid circle) and the crystal (open circle) as a function of the inverse available free volume. Maximum packing is 0.66 (RCP) for the glass and 0.74 for the fcc crystal. The solid line in figure 9b suggests that the initial increase of $\gamma_{cross}$ up to its maximum follows a power law behavior common for both the glass and the crystal.

Figure 10: Waiting time dependence of G' for glasses (solid symbols) and shear induced crystals (open symbols) at three different volume fractions for the small spheres in octadecene: φ=0.656 (squares), 0.641 (triangles) and 0.631 (circles) in a semi-logarithmic plot. The solid lines denote the logarithmic increase. Measurements were conducted every 30 minutes for a short time in order to minimize stresses in the sample.

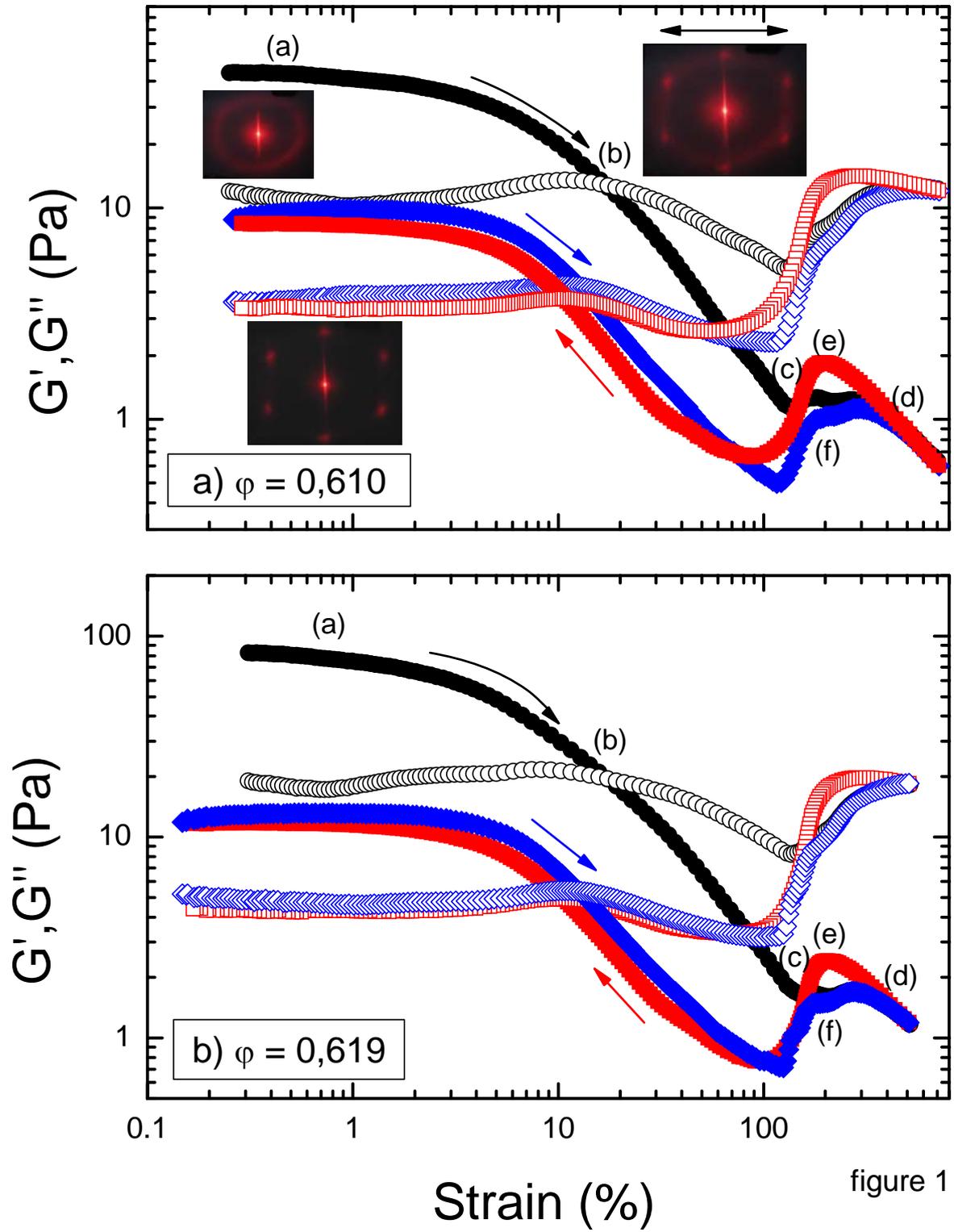

figure 1

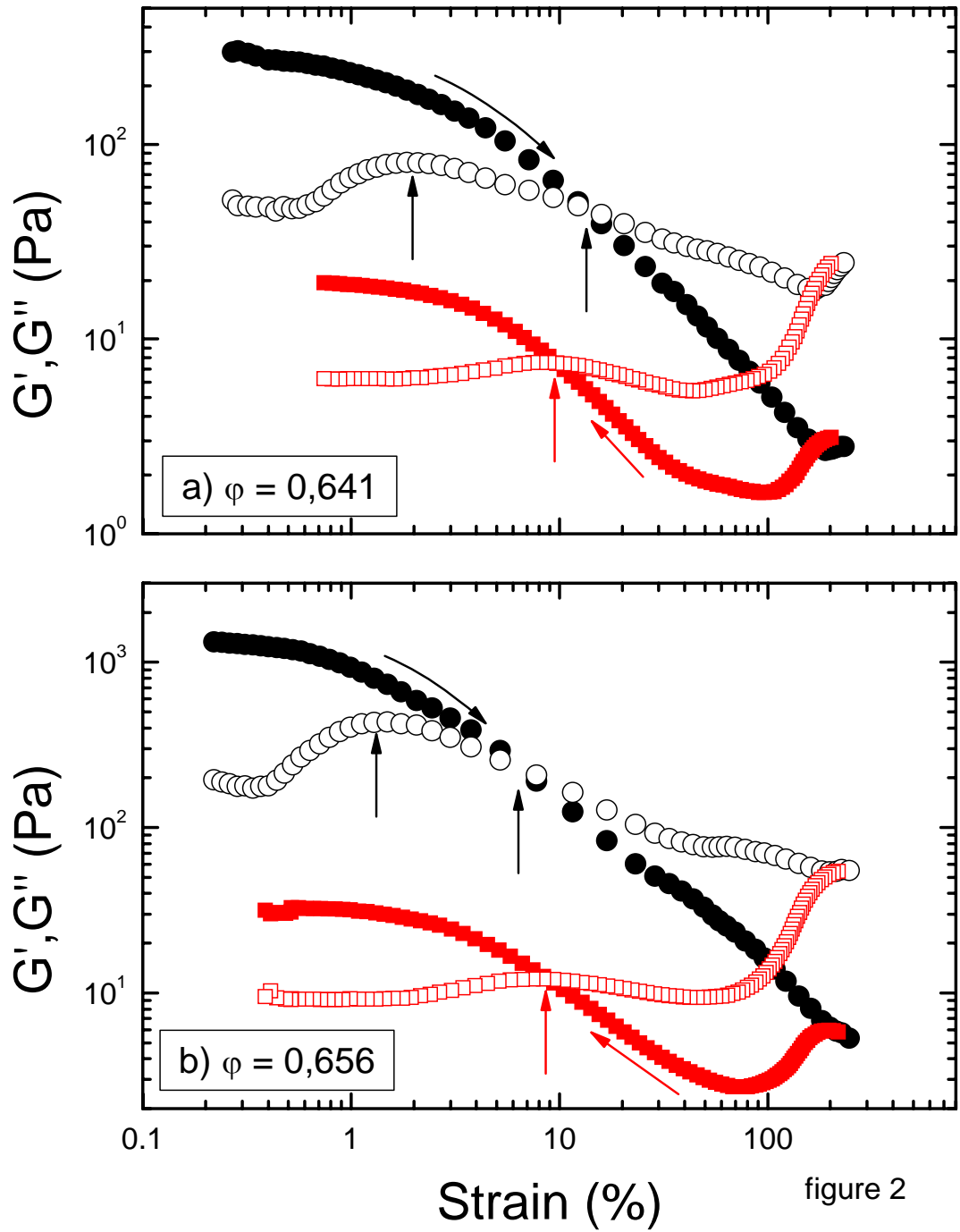

figure 2

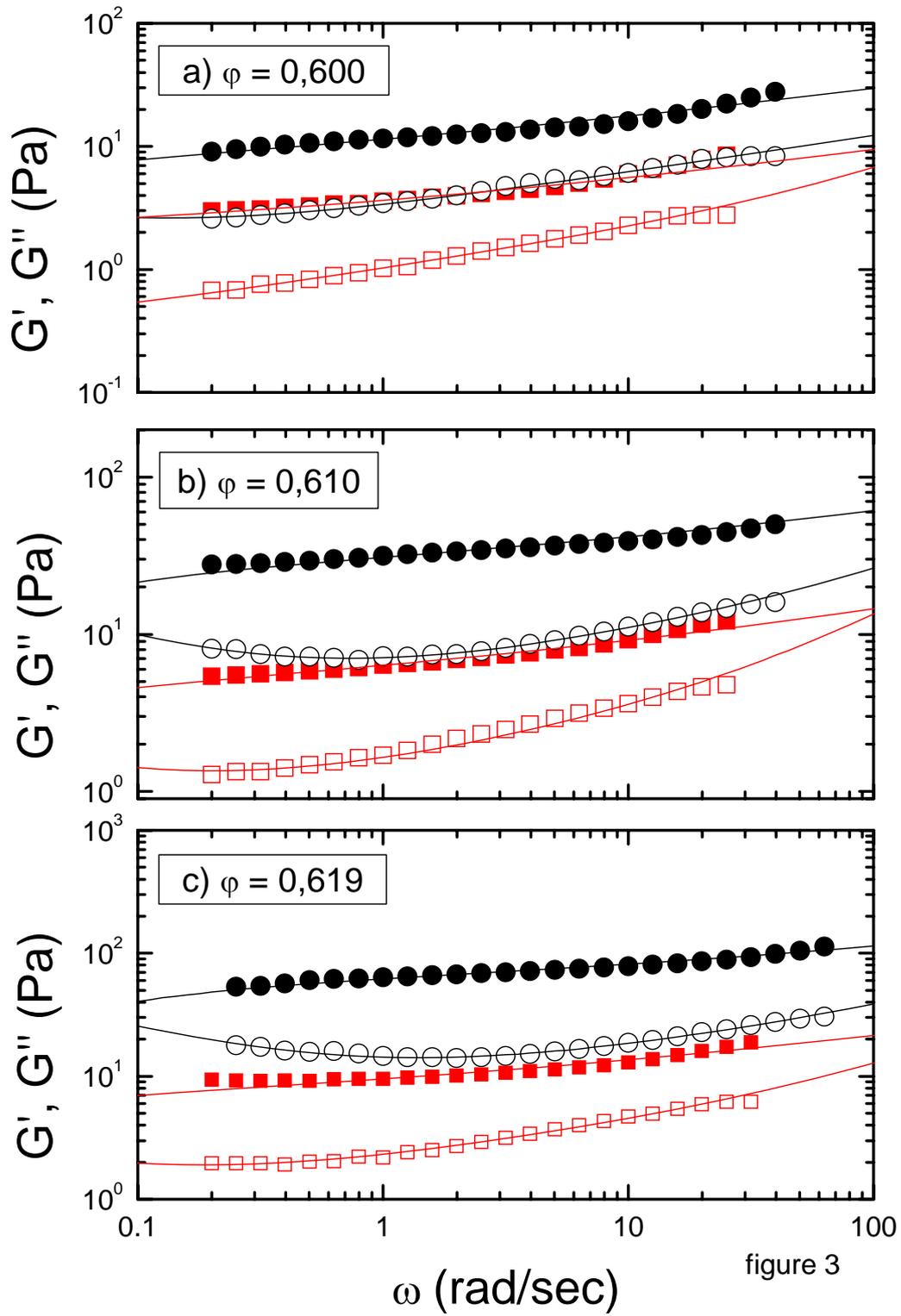

figure 3

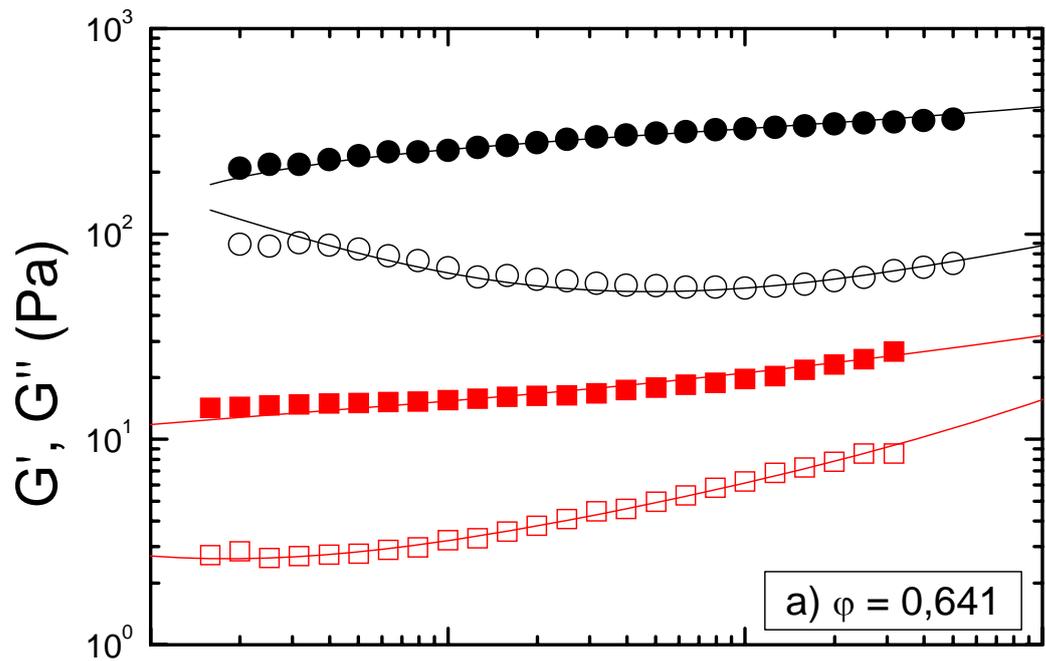
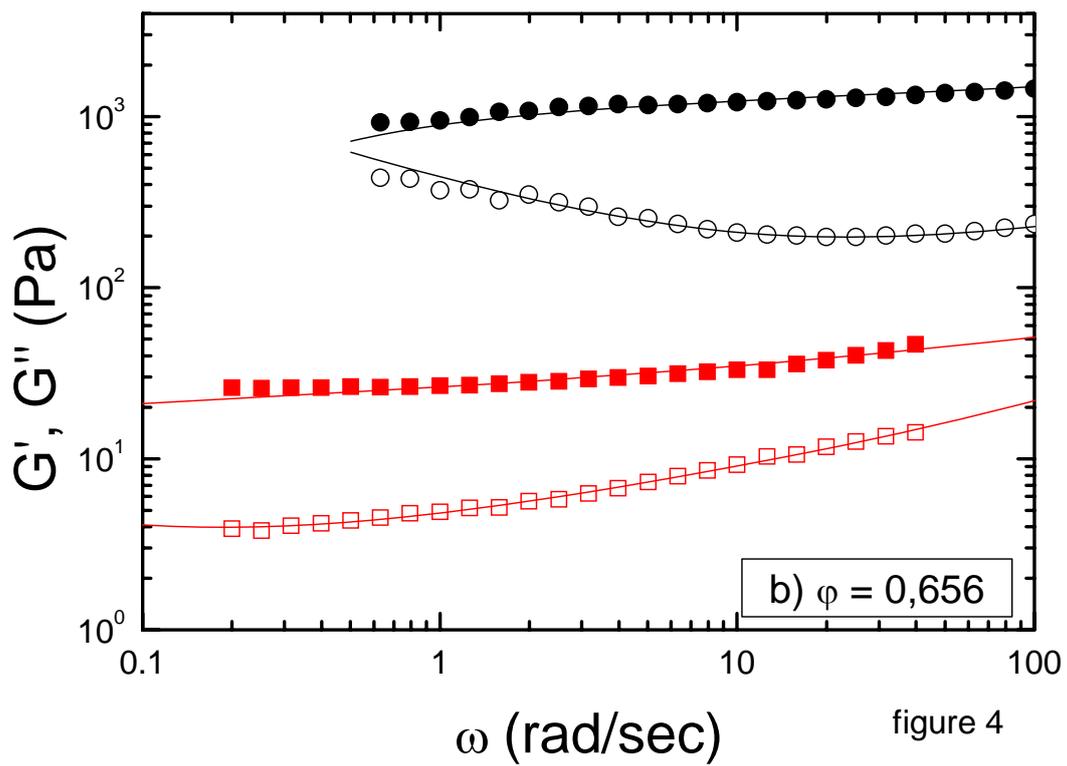

figure 4

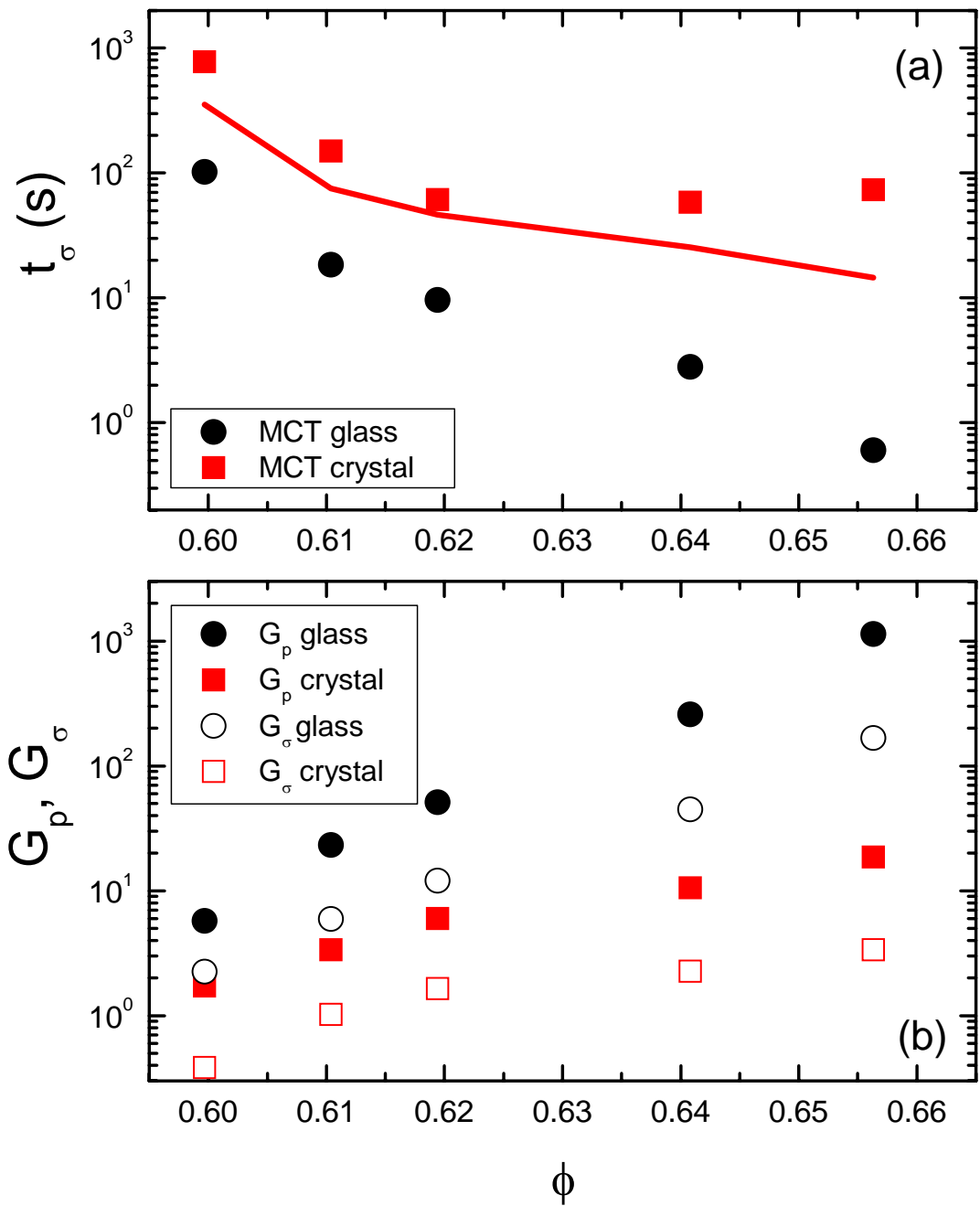

figure 5

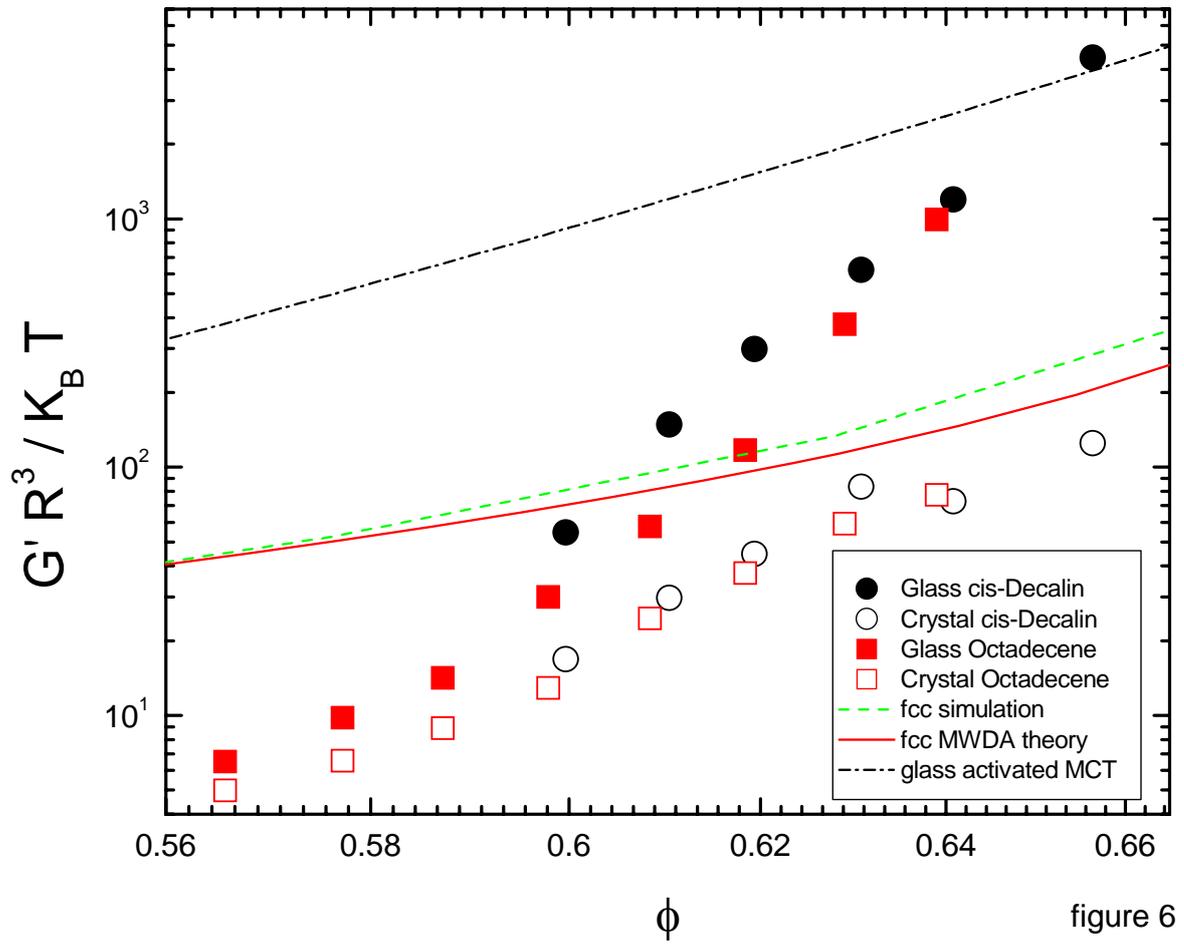

figure 6

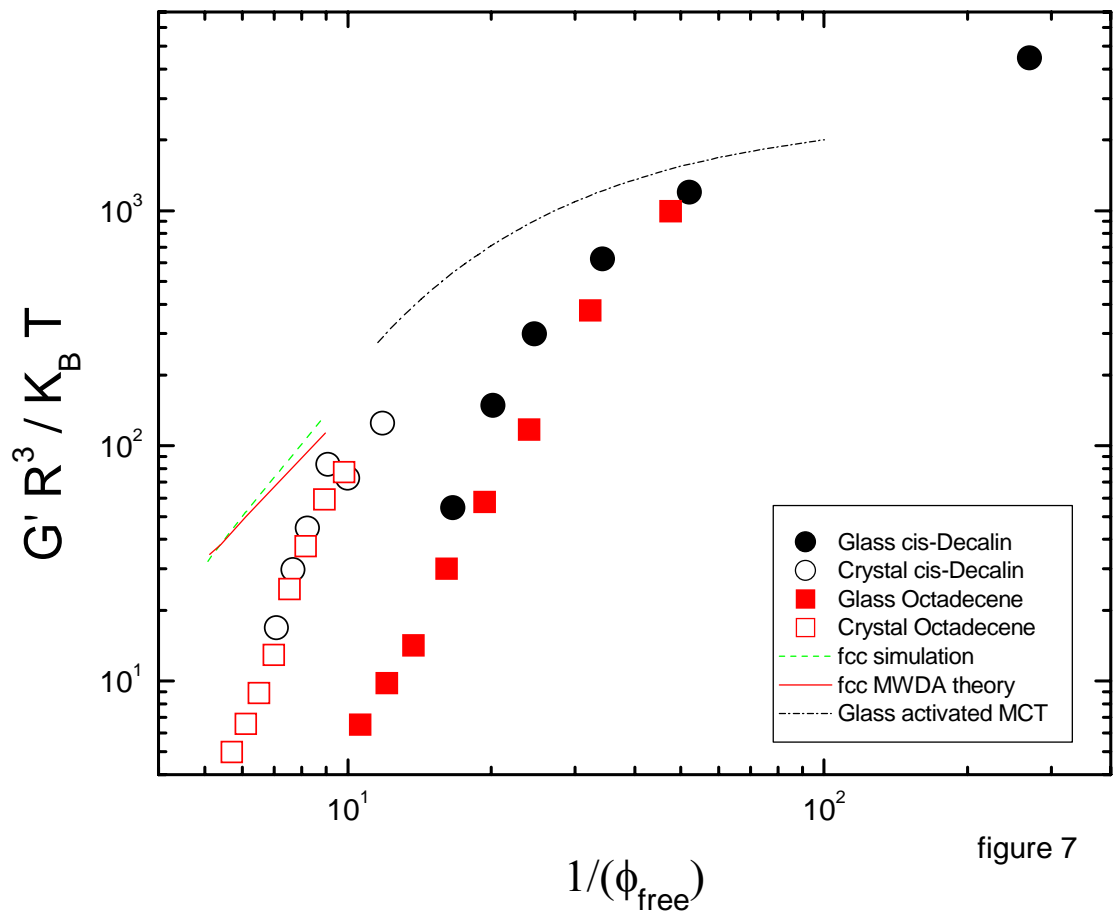

figure 7

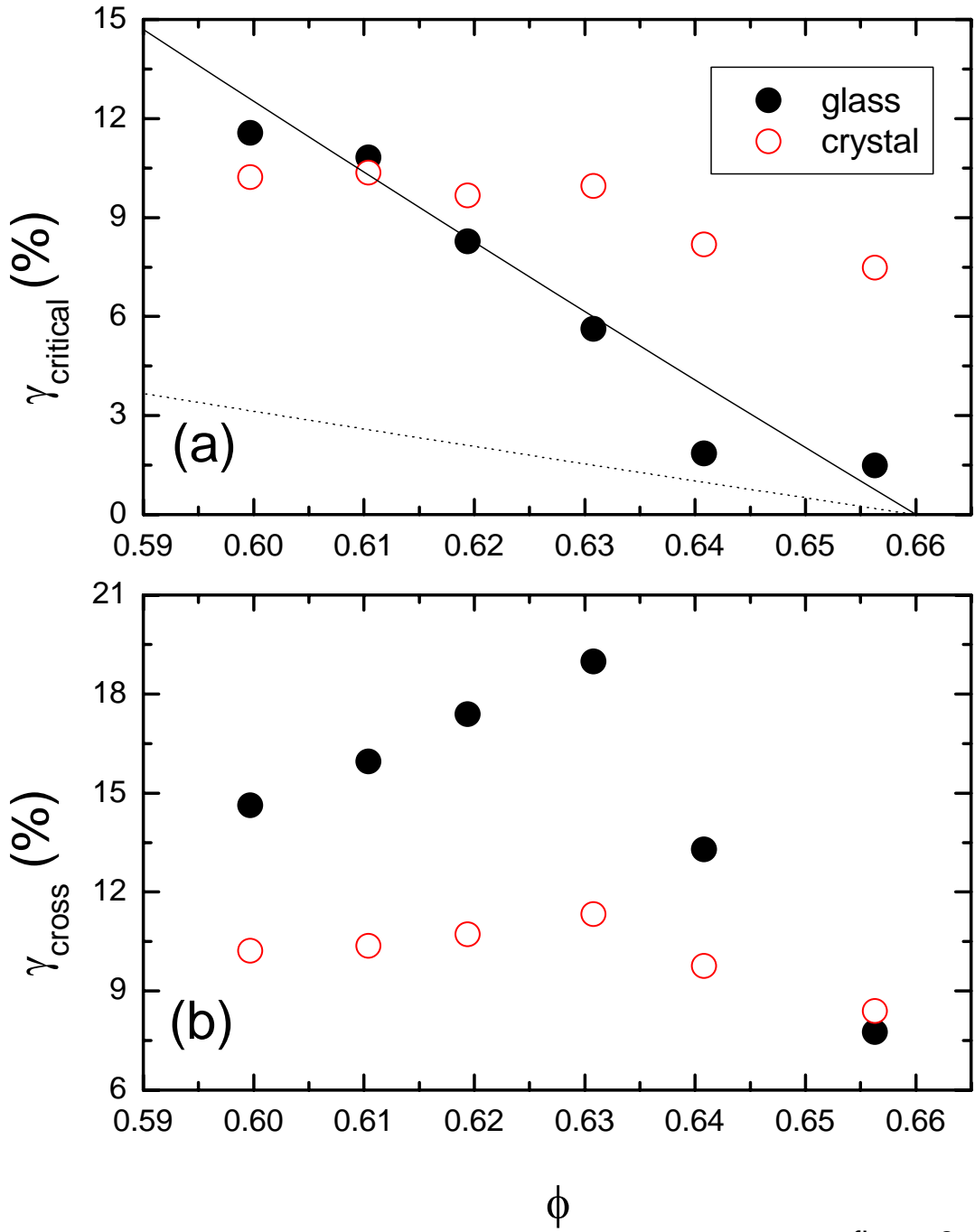

figure 8

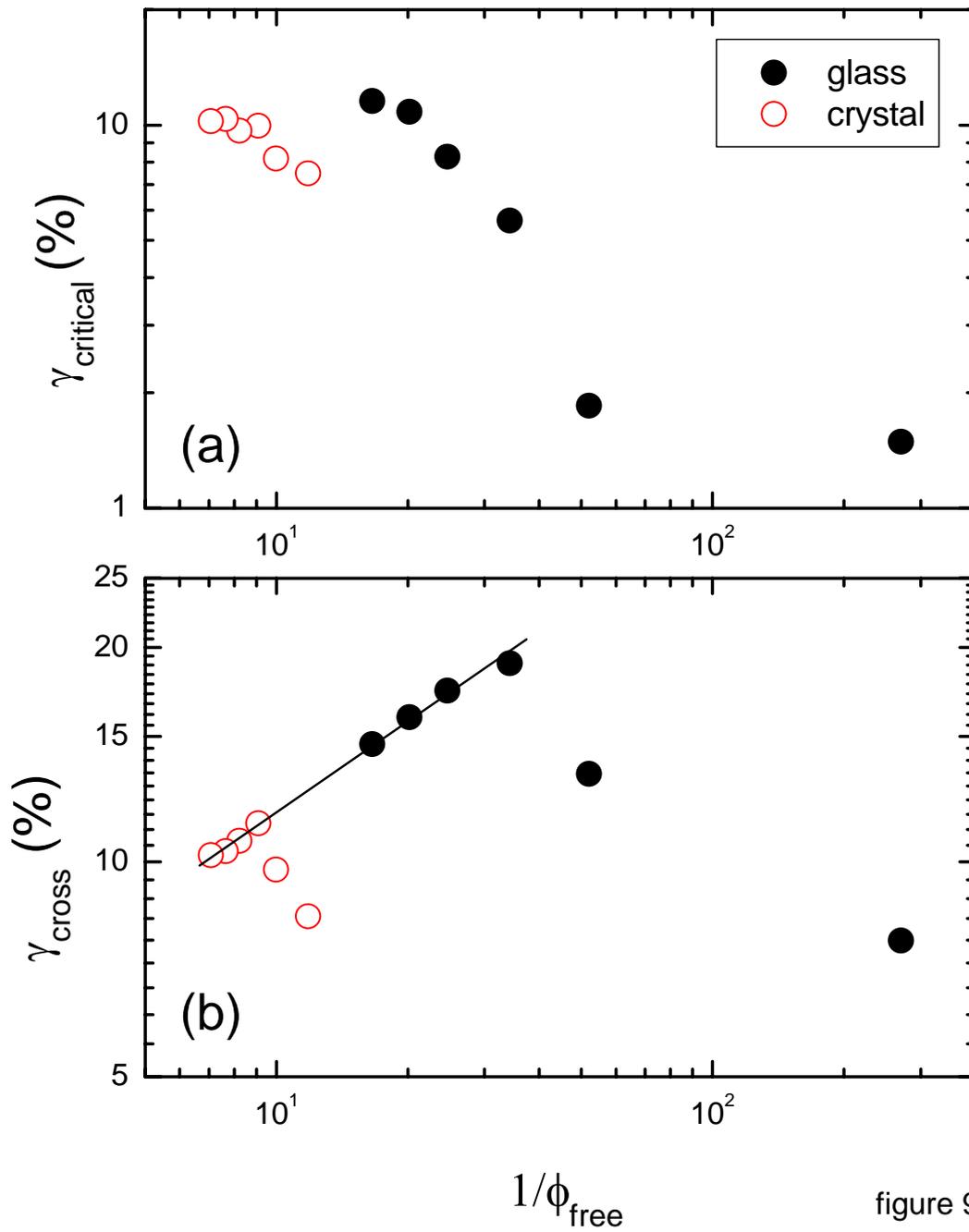

figure 9

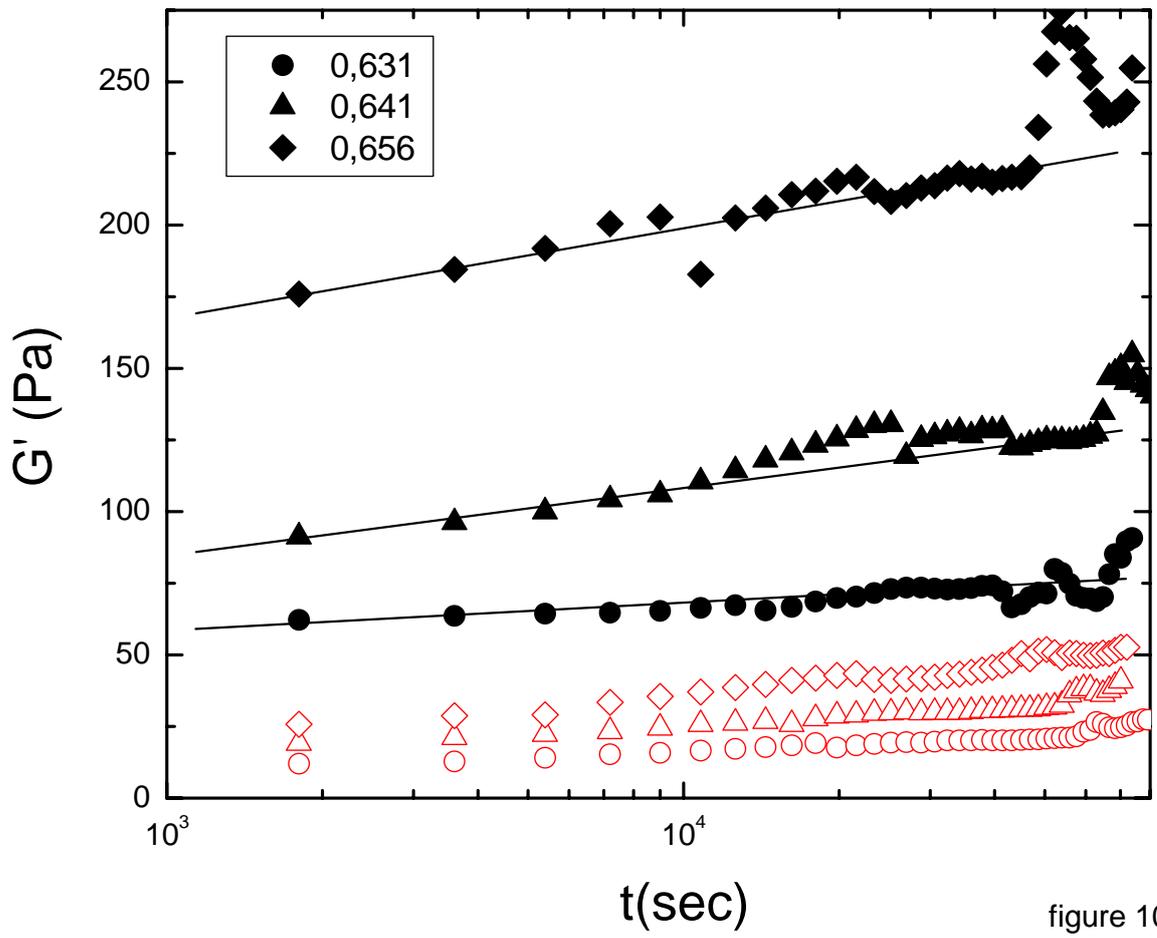

figure 10